\documentclass[12pt]{elsarticle}
\usepackage{epsfig}
\usepackage{amsmath,amsfonts,amssymb}

\begin{document}

\title{Heterogenous scaling in interevent time of on-line bookmarking}

\author{Peng Wang$^{1,2}$}
\ead{wangpenge@gmail.com}
\author{Xiao-Yi Xie$^1$}
\author{Chi Ho Yeung$^2$}
\author{Bing-Hong Wang$^{1,3}$}
\ead{bhwang@ustc.edu.cn}

\address{$^{1}$Department of Modern Physics, University of
Science and Technology of China, Hefei 230026,
China\\$^{2}$Department of Physics, University of Fribourg,
Fribourg, 1700, Switzerland
\\$^{3}$The Research
Center for Complex System Science, University of Shanghai for
Science and Technology and Shanghai Academy of System Science,
Shanghai, 200093 China}

\begin{abstract}
In this paper, we study the statistical properties of bookmarking
behaviors in Delicious.com. We find that the interevent time ($\tau$)
distributions of bookmarking decays powerlike as $\tau$ increases at
both individual and population level. Remarkably, we observe a
significant change in the exponent when interevent time increases
from intra-day to inter-day range. In addition, dependence of exponent
on individual $Activity$ is found to be different in the two ranges.
These results suggests that mechanisms driving human actions are different in intra-
and inter-day range. Instead of monotonically increasing with $Activity$, we find that
inter-day exponent peaks at value around 3. We further show that
less active users are more likely to resemble poisson process in
bookmarking. Based on the temporal-preference model, preliminary
explanations for this dependence have been given . Finally, a
universal behavior in inter-day scale is observed by considering the
rescaled variable $\tau / \langle \tau \rangle$.
\end{abstract}

\maketitle



\section{Introduction}
With increasing availability of data from internet applications,
recent years have witnessed expanding interest to characterize and
model human behaviors. Many on-line human activities such as email
communications \cite{dm7,dm8,dm1,dm2}, web surfing
\cite{dweb0,dweb1,dweb2}, movie rating \cite{activity1}, online game
\cite{dgame1}, blog posting \cite{dblog1} and off-line activities
such as letter communications \cite{dlett1,dm2,m4} and text message
\cite{dmessage1} are under active investigation which provide
understanding of our society. One of the main results from these
empirical studies is the heavy-tailed nature of the interevent time
distribution: the time interval between two consecutive human
actions, which we denote as $\tau$, follows a power-law
distribution, i.e. $P(\tau) \sim \tau^{-\beta}$. Moreover, some
studies claimed that there exist a few universality classes in human
dynamics characterized by universal exponents \cite{dm8}, which lead
to scientific debates \cite{dm1,dm2,activity1,dweb2}. Other studies
show that the exponents of inter-event time distribution depend on
$Activity$ (the frequency an individual takes actions), which
implies that exponent of individual is not a well representative for
human behaviors \cite{activity1,dweb2}, but an universal behavior
can be anyway found by considering the rescaled variable $\tau /
\langle \tau \rangle$ \cite{dm2,dweb2}. It is noted that this strong
dependence can only be observed in the inter-day range and it
becomes much weaker in the intra-day range \cite{dblog2}. These
results also suggest that we may classify human activities by
different time range.

In this paper, we study in details interevent time statistics in
Delicious.com, which is a typical web 2.0 application. Through
Delicious, users save and manage bookmarks, while sharing
interesting bookmarks with friends. It should be noted that there is
a close relation between web surfing and bookmarking: in most case,
we surf on web, bookmark interesting webpages, and continue surfing.
Heavy-tails were already observed in the distribution of time between consecutive visits to
URLs \cite{dweb1,dweb2}. On the other hand, the data set of
Delicious is widely adopted as training sets for recommender
systems \cite{recommender1, recommender2}. Understanding the
temporal pattern in Delicious may give insight to devise time-aware
recommender algorithms, which utilize time stamps of data to increase
recommendation accuracy \cite{recommender3, recommender4}.

The paper is organized as follows. In Section 2, we provide detailed
descriptions of the data set studied. In Section 3,  we give examples
of individual interevent time distributions which show heavy-tailed
nature and heterogenous scaling in intra-day and inter-day range. In
Section 4, we give the global interevent time distribution in these
two ranges and distinguish them by estimating the decay exponents
respectively. In Section 5, through comprehensive analysis of
exponent dependence on $Activity$, we show different trend observed
in the intra- and inter-day scale. Then a data collapse among the
inter-day distributions is observed by considering the rescaled
variable $\tau / \langle \tau \rangle$. Finally, we conclude and
summerize the results in Section 6.

\section{number distribution}
Our data set consists of 54204641 bookmarking activities by 220867
users over a period of 31 months (between 2004/04/01 and
2007/11/01). Here we use only the identifier (ID) of the users and
the time when the bookmarks were saved. The resolution of time
stamps is in seconds. Figure 1 shows the distribution of $k$, the number
of bookmarks saved by a single user. As we can see, $P(k)$ is broad
and the tail of the distribution decays as power-law as $k$
increases, giving $P(k) \sim k^{-2.41}$. This result resembles the
distribution of messages number in Ebay \cite{dweb2},  and is
significantly different from that of logging action in wikipedia
(which follows power-law over the whole range \cite{dweb2}) and post
number in blog (which is so called ``double power
law" \cite{dblog1,dblog3}). Interestingly, in spite of the difference
in these distributions, statistics on interevent times of them are
very similar as we will see below.

\begin{figure}[htb]
\centering
\includegraphics[width=0.6\textwidth]{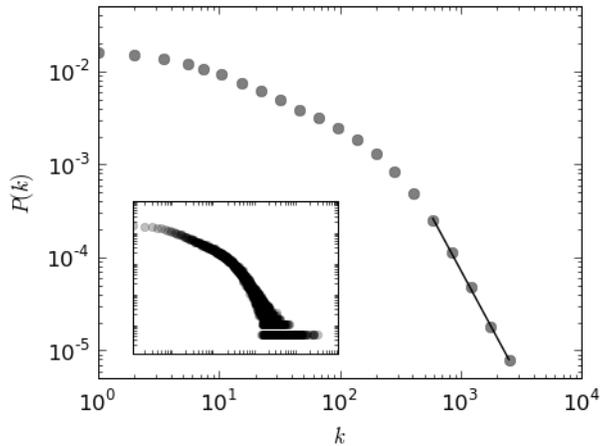}
\caption{\label{Fig:RI:PDF}The distribution of bookmark number $k$
of user in Delicious. The decay exponent is 2.41.}
\end{figure}

\section{Interevent time distribution for individuals}

In our context, the interevent time $\tau$ is defined as the time
interval between consecutive bookmarks by the same user. Figure 2
shows the cumulative distribution of interevent time obtained from
six users. As we can see, all curves show a crossover around $\tau
\approx 1$ day, which correspond to a change in exponent between intra- and inter-day
range. Although power-law decays are observed in both ranges, the
change in exponent (which is also noticed in other
systems \cite{dblog2}) suggests that the mechanism driving intra- and
inter-day activities are different. Moreover, changes in exponent
are observed even within the intra-day range for some users. As shown
in figs 2e and 2f, a slight increase in the decay exponent is
observed at $\tau \approx 1$ hour.

\begin{figure}[htb]
\centering
\includegraphics[width=1.\textwidth]{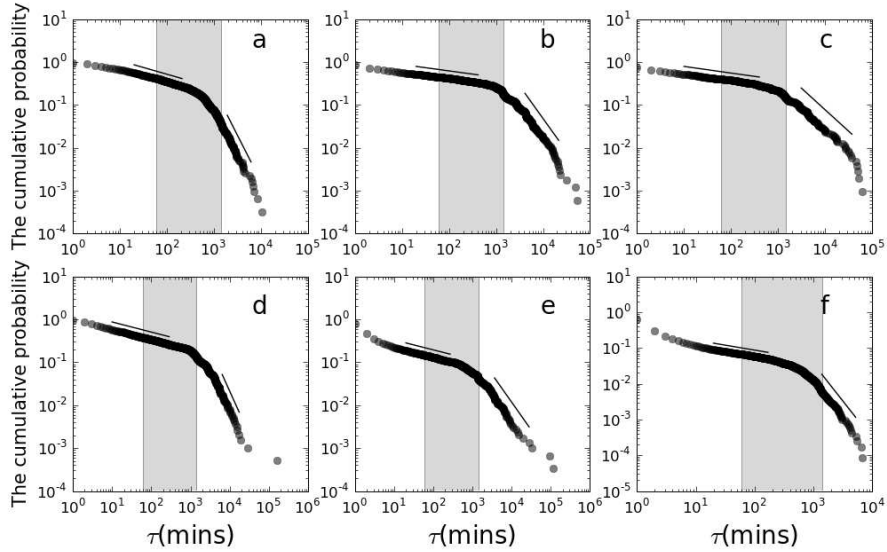}
\caption{\label{Fig:RI:PDF}The cumulative distribution of interevent
time of six random individuals. Their corresponding number of
bookmarks are 3104, 1689, 1047, 1946, 2983,11892. The shaded area
correspond to the span of 60 mins (1 hour)  $< \tau <$ 1440 mins (1
day). The exponents of these cumulative distributions in the intra-
and inter-day range ($\beta_{intra}, \beta_{inter}$) are: (a)(0.31,
2.15); (b)(0.15, 1.53); (c)(0.15, 1.0);(d)(0.23, 2.02); (e)(0.23,
1.29); (f)(0.28, 2.09).}
\end{figure}

\section{The global distribution of interevent time}

\begin{figure}[htb]
\centering
\includegraphics[width=1.0\textwidth]{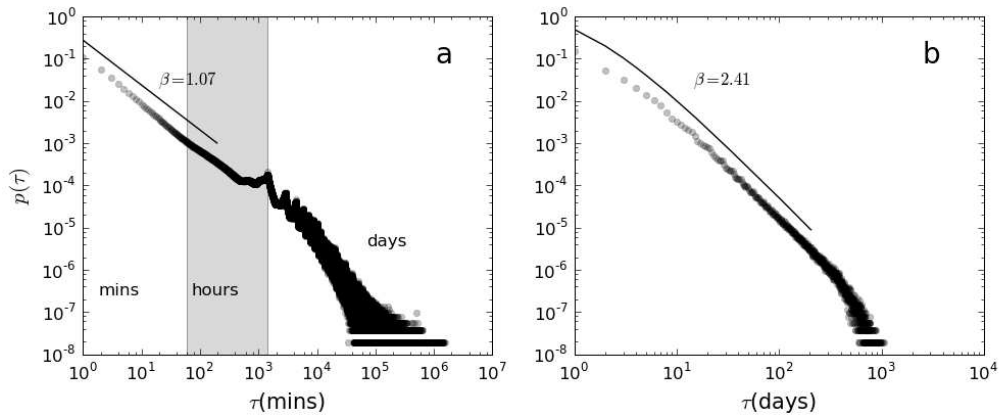}
\caption{\label{Fig:RI:PDF}The global distribution of interevent
time with precision in (a) one minute and (b) one day.}
\end{figure}

The global distribution of interevent time are plotted in fig. 3. In
order to have a clear picture in the intra-day range, we express
$\tau$ in fig. 3a with a resolution of minutes. In fig. 3b, we express
it with a resolution of days where the circadian oscillation are
masked which makes the decay in the inter-day range clearer. Both
distributions in the intra- and inter-day range present a powerlike
decay with exponents $\beta_{inter} \simeq 1.07$ for intra-day range
and $\beta_{intra} \simeq 2.41$ for inter-day range. This significant
difference between the exponents of the two ranges is consistent with
the results obtained from the distribution of individuals in Sec 3. The
exponent in the intra-day distribution in this case is the same as
the one obtained from consecutive visits of URL \cite{dweb1}. It is
reasonable, since bookmarking often follows web surfing as we
mentioned above. The exponent of the inter-day distribution is
very large compared to other systems \cite{activity1,dblog2,dweb2}
as we know that would make some difference in the following analyses. It should
be noted that we fit the inter-day distributions with the so-called ``shifted
power-law" \cite{book}:
\begin{equation}
y\sim(x+h)^{-\beta}
\end{equation}
When $h$ is large this distribution tends to resemble exponential
distribution, and it approaches a power law when $h \rightarrow 0$ \cite{book}.
As we know, human activities based on poisson
process (where event occurrence are independent and have identical
probability) can lead to an exponential distribution of interevent
time. Therefore, $h$ somehow measures the extent that human activity
resembles poisson process, i.e. homogeneous probability to bookmark
webpages over time. For instance, the distribution in fig. 3b is
fitted by eq. (1) with $h \approx 3.32$. We will further discuss the
fitting values of $h$ for distributions with different individual
$Activity$ in the next section.

\section{Activity and exponents}
Based on the heterogeneity, we perform investigation on both the
intra- and inter-day distributions. Firstly, we define the
average $Activity$ $A_{i}$ of user $i$ as
\begin{equation}
A_{i}=\frac{n_i}{T_i}
\end{equation}
where $n_i$ is the total number of bookmarks saved by user $i$ and
$T_i$ is the time interval between the first and the last bookmark
of user $i$. We consider only users with at least 20 bookmarks and
$T_i$ which is more than 10. There are 173108 users who satisfy
these conditions. Figure 4 shows the distribution of $Activity$
$A_{i}$ which approximately follow a log-normal distribution, as
shown by the fitting line. As we can see, the $A_{i}$ of most user
is between 0.01 and 1 with most probable value around 0.2 per day.

\begin{figure}[htb]
\centering
\includegraphics[width=0.6\textwidth]{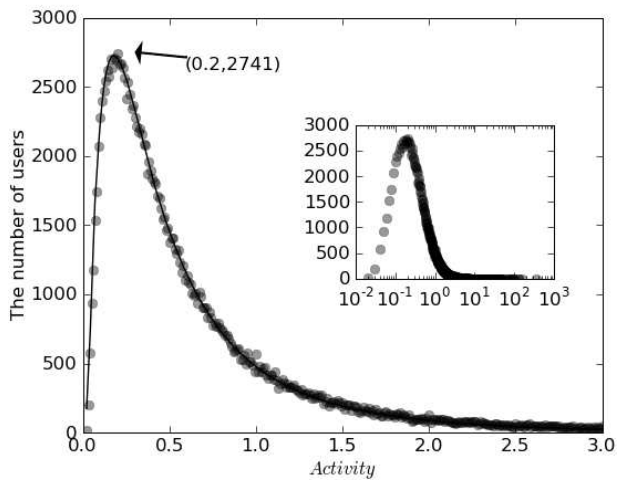}
\caption{\label{Fig:RI:PDF}The distribution of $Activity$. The solid
line corresponds to the fitting of log-normal distribution $\ln
{\cal N}(\mu, \sigma^2)$ with $\mu = \ln(0.42)$ and $\sigma=0.93$.}
\end{figure}

To examine interevent time distribution in relation with user
$Activity$, we then sort users in an ascending order of $A_i$ and
divide the entire population into 10 groups, each of which have $M$
users ($M \approx N/10$ where $N$ is the total number of users).
Accordingly, the mean activity of each group obeys the inequality
$\langle A \rangle_1 < \langle A \rangle_2 ... < \langle A
\rangle_{10}$. We then investigate the dependence of exponent on
$Activity$ in both the intra- and inter-day range. In fig. 5, we
plot the interevent time distribution of group 1, 5 and 9 (which
respectively correspond to $\langle A \rangle=0.09, 0.37, 1.12$ per
day). In the intra-day range, we find that the exponents only weakly
dependent on $\langle A \rangle$ (as shown by fig. 5a a slight
decrease is observed with $\langle A \rangle$). On the contrary, in
the inter-day range, the exponents increase from 2.15 to 2.91 with
$\langle A \rangle$ from 0.37 to 1.12. This result shows that
behavioral heterogeneity in intra- and inter-day range is also
evident from exponent dependence on $\langle A \rangle$.

\begin{figure}[htb]
\centering
\includegraphics[width=1.0\textwidth]{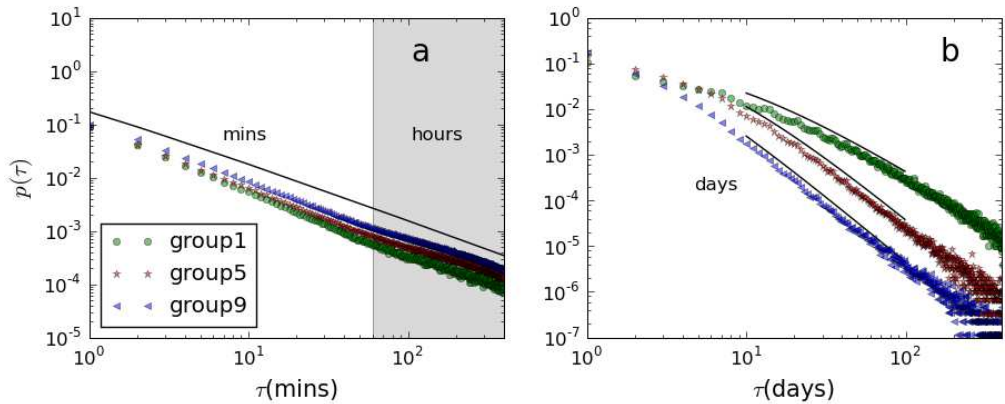}
\caption{\label{Fig:RI:PDF}(color online) The exponent dependence
with $Activity$. The interevet time distributions of group 1, group
5 and group 9 are shown in this figure. As a comparison, the slope
of the straight line in fig a is 1.07 which we got from the global
distribution. In fig b, the exponents we got are 2.15 for group 1,
2.91 for group 5 and 2.90 for group 9.}
\end{figure}

In fig. 6, we show the inter-day exponent of the interevent time distribution
 as a function of $A_i$ in fig 6. It is noted that
the exponents here increase much faster than the ones in previous
studies in spite of the same tendency \cite{activity1,dweb2,dblog2}.
The exponent of group 1 is $2.15$ and the one of group 3 already
increase to 2.74. The reason for this steep increase has to do with the large exponent of
the global distribution which is 2.41 as we mentioned above. The
advantage is that it leads us to observe this dependence on a
broader range. As we can see, the exponent reaches the maximum at
group 6 and then it decreased slightly. It suggests there is a limiting
value ($\beta \simeq 3$) of decay exponents. Actually, in Radicchi's
study\cite{dweb2}, the exponents of last group of America On-Line and
Wikipedia also decrease. We
further plot in the inset of fig. 6 the fitted values of $h$ from eq. (1) for the
distributions of the 10 groups, which shows a monotonic decrease of
$h$ with $\langle A \rangle$. For instance, $h \approx 7$ for group
1 and  $h \approx 0.7$ for group 10, indicating a substantial
decrease of $h$. As mentioned before, it shows that poisson process plays
a more significant role in the behavior pattern of less
active users. We can understand these results based on the
temporal-preference model which has two ``choose rule" \cite{dblog1}:
(1) the more the user performs an activity recently, the more likely
he will do it next; (2) there exists occasions that user choose what
to do randomly with independent probability. In relevance to the model, the behaviors of
the inactive users are relatively more likely to follow the random
rule. Similar dependence of exponents on $\langle A \rangle$ is actually observed
in this model \cite{dblog1}.

\begin{figure}[htb]
\centering
\includegraphics[width=0.6\textwidth]{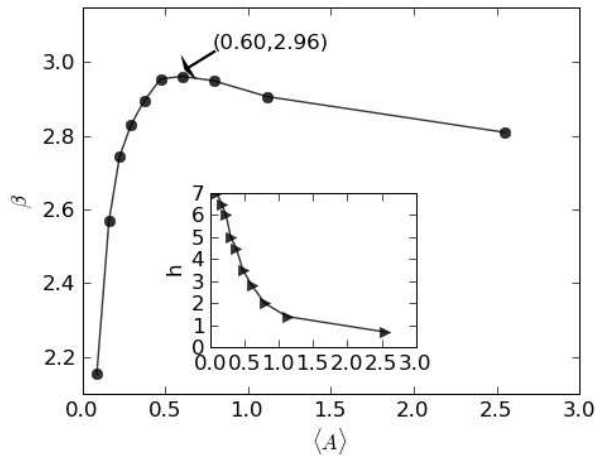}
\caption{\label{Fig:RI:PDF}$\beta$ of each group is plotted as a
function of average $Activity$. The one of $h$ is also plotted in
the inset.}
\end{figure}

In fig. 7, these distributions are rescaled with the average interevent
time $\langle \tau \rangle$ of the respective group of users.
As we can see, the scaling produces a data collapse between the different
curves. This observations, which are already noticed in other
systems\cite{correlation1,dweb2,mobile1}, implies that the mechanism
underlying individual behaviors within the inter-day range is still the
same in spite of different exponents. Furthermore, the $Activity$ of
individual users play an important role in this mechanism.

\begin{figure}[htb]
\centering
\includegraphics[width=0.6\textwidth]{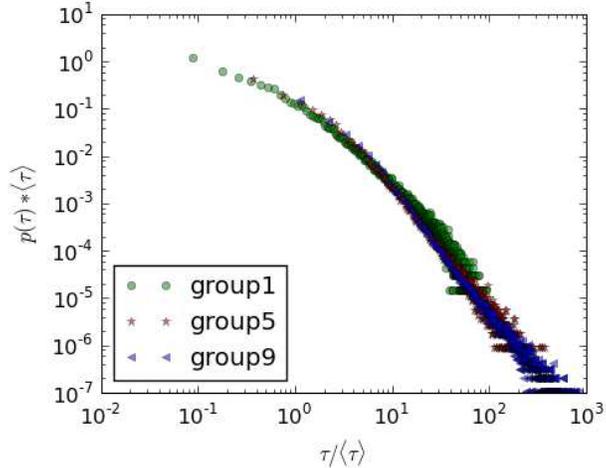}
\caption{\label{Fig:RI:PDF}(color online) Scaling of the interevent
time distributions.}
\end{figure}

\section{conclusion}
In this paper, we show the heavy-tailed nature of the distribution of
interevent times at both individual and population level. Our
results further verified heterogenous human dynamics in intra- and
inter-day ranges. On one hand, there is a significant difference
between the exponents in the intra- and inter-day distribution,
which are 1.07 and  2.41. On the other hand, the inter-day exponents are
strongly dependent on individual $Activity$, while this dependence is absence in intra-day
range. Moreover, our study suggests that there is a maximum value of
$\beta \approx 3$ for the increase of exponent with $Activity$. The
distributions in intra-day range seem to be in universality classes of
human dynamics characterized by exponents of $\beta =
1$ \cite{dblog1,dweb1,dweb2,dm8,m4}, if we ignore the weak dependence
of exponents on $Activity$. There are already models
which produce the interevent time distribution with similar
exponents \cite{m2,dblog21}. The strong dependence of exponent on
$Activity$ in the inter-day range implies that it is inappropriate
to classify the distributions in this range by only their exponents.
However the universality still exists in inter-day human dynamic. One
evidence is the similar exponent dependence on $Activity$ which are
observed in many systems \cite{activity1,dweb2,dblog2}. The data collapse
between these different distributions provide further confirmation. In
general, the exponents of inter-day distribution are often greater
than that of intra-day distribution and smaller than
3 \cite{activity1,dweb2,dblog2}. The temporal-preference model, which
is built to simulate the behavior pattern of blog-posting, can already
give a preliminary explanation for the inter-day dynamics. However
the current model is oversimplified. To improve and give better agreements with most systems, it
is the key to understand the relation between action repetition and
memory\cite{Repetitions1,Repetitions2}.

\section*{ACKNOWLEDGMENTS}
This work is funded by the National Basic Research Program of China
(973 Program No.2006CB705500); The National Important Research
Project:(Study on emergency management for non-conventional happened
thunderbolts £¬Grant No. 91024026); the National Natural Science
Foundation of China (Grant Nos. 10975126£¬10635040),the Specialized
Research Fund for the Doctoral Program of Higher Education of
China(Grant No.£º20093402110032)

\end{document}